\documentclass[11pt]{article}
\usepackage{graphicx}
\begin{document}

\title{An exact spin liquid state on the kagome lattice with topological order}

\author{Zhong Wang and Shaolong Wan\thanks{Corresponding author.
Electronic address: slwan@ustc.edu.cn} \\
Institute for Theoretical Physics and Department of Modern Physics \\
University of Science and Technology of China, Hefei, 230026, {\bf
P. R. China}}

\maketitle
\begin{center}
\begin{minipage}{120mm}
\vskip 0.8in
\begin{center}{\bf Abstract} \end{center}

{We find an exact spin liquid state without time reversal symmetry
on the kagome lattice with odd number of electrons per unit cell and
explicit wave functions for all eigenstates. We also obtain that all
spin-spin correlations are zero except trivial cases. We then show
that there are anyonic excitations in our model. Finally, we
demonstrate the existence of ground state degeneracy and gapless
edge states indicating nontrivial topological order in our model. We
also label all eiginstates on torus by local and global string
operators.}

\end{minipage}
\end{center}

\vskip 1cm

\textbf{PACS} number(s): 75.10.Jm, 75.50.Mm

\emph{Introduction.} --There has been enormous interest recently in
topological quantum phases \cite{wen1,nayak} beyond the Landau
paradigm \cite{sachdev}, which is based on the idea of symmetry
breaking and local order parameters. The newly discovered phases
have many exotic phenomena such as electron fractionalization
\cite{laughlin} and emergent gauge fields \cite{anderson}. The
traditional order parameter descriptions are either inapplicable or
insufficient for these new phases.

Spin liquid states in dimension two ($2$D) are examples of such
exotic phases. In the spin liquid states, strong quantum fluctuation
forbids existence of nonzero spin order parameter. Although many
numerical and analytical evidences indicate the existence of such
exotic state in $2$D, exact solution is absent until very recently
\cite{kitaev1,kitaev2,wen2,yao,schroeter,sun,xiang}. These exact
solutions provide us with a lot of intuition of the internal
structure of spin liquid states.

Spin liquid states that spontaneously break time reversal symmetry
(TRS) and spacial inversion symmetry are called chiral spin states
\cite{kalmeyer}. An exact chiral spin state is proposed in
Ref.\cite{yao} based on short-range anisotropic interaction
\cite{kitaev2}, but this model have even number of electrons per
unit cell, and thus not deserve to be called ``Mott insulators'' or
``spin liquid'' states in the traditional sense. We consider it an
interesting question to find whether there exist other exact spin
liquid states with arbitrarily given symmetry properties and odd
number of electrons per unit cell.

In this letter we generalize the toric code model
\cite{wen2,kitaev1} to the kagome lattice (Fig.1) and propose an
exact spin liquid state that breaks TRS. This state is not chiral
spin state in conventional sense because TRS is already broken in
the Hamiltonian. The kagome lattice has odd number of spins per unit
cell, and thus the proposed state is a true spin liquid state. Our
model treat three spin components $\sigma^{x,y,z}$ democratically,
which is more symmetrical than the $\sigma^{x,y}$ model
\cite{wen2,kitaev1}. We propose explicit wave functions for all the
energy eigenstates and interpret these eigenstates as string-net
\cite{wen1} coherent states. We also compute spin-spin correlation
functions and find that all of them are zero except the trivial
cases of on-site correlations. After a short discussion of anyonic
excitation in our model, we proceed to discuss the ground state
degeneracy and gapless edge states indicating topological order. We
also find all the eigenstates of the model on torus using string
operators.

\begin{figure}
\includegraphics[width=8.5cm, height=6.0cm]{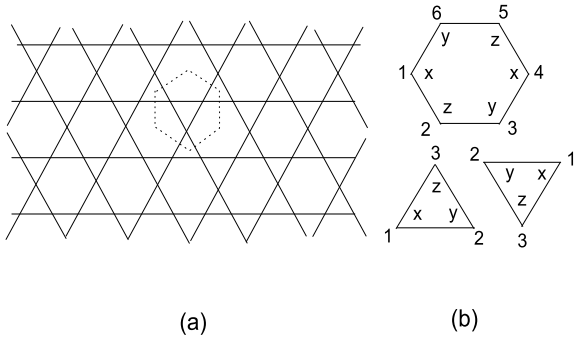}
\caption{(a) The kagome lattice is composed of two types of
elementary plaquettes: triangles and hexagons. The dashed lines show
the unit cell. (b) Elementary plaquettes of Kagome lattice. The
letters x,y,z indicate the factors of $F_{p}$.} \label{fig.1}
\end{figure}

\emph{The model}. --Our spin-$\frac{1}{2}$ Hamiltonian on the kagome
lattice is given as:
\begin{eqnarray} H = -\sum_{p}V_{p}F_{p}
\end{eqnarray}
with summation over all elementary triangular and hexagonal
plaquettes. We choose $F_{p}=\sigma_{1}^{x} \sigma_{2}^{y}
\sigma_{3}^{z}$ on the triangles and $ F_{p}=\sigma_{1}^{x}
\sigma_{2}^{z} \sigma_{3}^{y}\sigma_{4}^{x} \sigma_{5}^{z}
\sigma_{6}^{y}$ on the hexagons (see Fig.1), where $\sigma_{i}^{a} $
are the well-known Pauli matrices. It follows from the property of
Pauli matrices that $F_{p}^2=1$. Another crucial property of $F_{p}$
is that $[F_{p},F_{q}]=0$ for any pairs $p$ and $q$. So all $F_{p}$
can have eigenvalues($\pm1$) simultaneously. Since $\sigma_{i}^{a}$
change sign under time reversal transformation, the Hamiltonian
breaks TRS explicitly. The solubility of our model does not depend
on signs of $V_{p}$, so we will choose all $V_{p}=V>0$ for
simplicity.

Firstly we consider a planar region with $N$ sites, where $N$ is
very large. The number of elementary triangle and hexagon are shown
to be $2N/3$ and $N/3$, respectively. So there are $N$ plaquettes,
and $F_{p}$ can label $2^{N}$ states. The number of physical spin
states is also $2^{N}$. From this counting argument we know that
eigenvalues of $F_{p}$ label all the states on an infinite plane and
the model is exactly solvable.

We can write down the following wave unctions for all eigenstates:

\begin{eqnarray}
|\psi_{\lambda}\rangle =
\prod_{p}(1+\lambda_{p}F_{p})|\psi_{0}\rangle
\end{eqnarray}
where $p$ exhaust all elementary triangular and hexagonal
plaquettes. $|\psi_{0}\rangle$ is an arbitrary spin state which we
choose as the state satisfying $\sigma_{i}^{z}=+1$ for all sites.
The irrelevant normalization factors have been omitted. Different
choices of $\lambda_{p}=\pm1$ label different states. The $F_{p}$
expectation in $|\psi_{\lambda}\rangle$ is:

\begin{eqnarray}
\frac{\langle\psi_{\lambda}|F_{p}|\psi_{\lambda}\rangle}{\langle\psi_{\lambda}|\psi_{\lambda}\rangle}
= \lambda_{p}
\end{eqnarray}
so $|\psi_{\lambda}\rangle$ is an eigenstate of $H$ with eigenvalue
$E = -\sum_{p}\lambda_{p}V_{p}$. The ground state is the state with
all $\lambda_{p}=1$. The above simple wave functions have
transparent explanation in the physical language of string-net
condensation \cite{wen1}. We will demonstrate this in our model. A
caution in order is that this wave function can completely determine
the state only on an infinite plane. On systems with boundary like
disk or closed like torus, there are other degrees of freedom
missing. We will return to this question later.

We define two types of strings $T_{1}$ and $T_{2}$
\cite{kitaev1,wen1}. $T_{1}$ strings walk on triangles while $T_{2}$
strings walk on hexagon (see Fig.2). We define the string operators
as $P_{t} = \prod_{i \in t}\sigma_{i}^{a} $ for each string $t$. The
choice of $\sigma_{i}^{a}$ depends on both the site $i$ and the
string type. For $T_{1}$ strings, $\sigma^{a} =
\sigma^{y},\sigma^{z},\sigma^{x}$ at the three vertexes $1$,$2$,$3$
of each triangle, respectively. For $T_{2}$ strings, $\sigma^{a} =
\sigma^{z},\sigma^{y},\sigma^{x},\sigma^{z},\sigma^{y},\sigma^{x}$
at the six vertexes $1,2,3,4,5,6$ of the hexagon, respectively. An
important property of this definition is that
 $T_{1}$ and $T_{2}$ string operators involve different Pauli matrices at any sites
 $i$, from which we can readily obtain the simple commutative
 relations among $P_{t}$. For arbitrary $T_{1}$ strings $s,t$ and $T_{2}$
strings $u,v$, we have $[P_{s},P_{t}]=0$, $[P_{u}, P_{v}]=0$;
$[P_{s},P_{u}]=0$ if $s$ and $u$ intersect each other by even times,
while $\{P_{s},P_{u}\}=0$ if $s$ and $u$ intersect each other by odd
times.

\begin{figure}
\includegraphics[width=8.5cm, height=6.0cm]{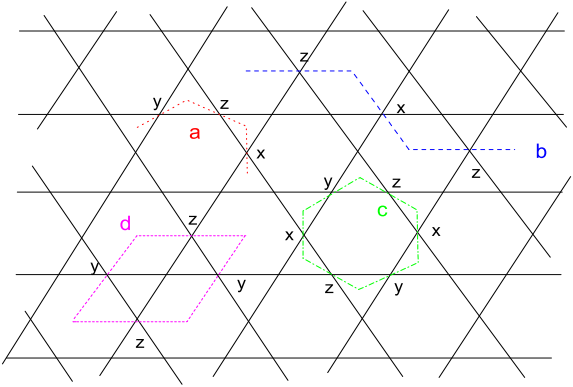}
\caption{Strings on the kagome lattice. $a,b$ are open strings,
while $c,d$ are closed strings. $a,c$ are $T_{1}$ strings, while
$b,d$ are $T_{2}$ strings. The letters $x,y,z$ denote the specific
Pauli matrices appearing in the string operators.} \label{fig.2}
\end{figure}

These strings are divided into two classes by their topology. A
string is called open string if it has endpoints, otherwise is
called closed string \cite{wen1}. From the property of Pauli
matrices we can find $[P_{t},F_{p}]=0$ for any $p$ for a closed
string $t$, and it follows from this that $[P_{t},H]=0$. For open
string $s$ with two endpoints on plaquettes $p,q$, we have
$\{P_{s},F_{a}\}=0$ if $a=p$ or $q$, and $[P_{s},F_{a}]=0$
otherwise. So $P_{s}$ changes the sign of $F_{p}$ at its two
endpoints. If we regard negative $F_{p}$ as vortex at $p$, then
$P_{s}$ applied to the ground state will creates two vortexes at its
endpoints and increase the energy by $2V$.

It is interesting that $F_{p}$ are also closed strings operators
defined above.  Actually, for a $ T_{1}$ string $t_{1}$ surrounding
a single hexagon $p$ (such as string $c$ in fig. 2), we have
$P_{t_{1}}=F_{p}$. Similar relations also hold for any $T_{2}$
string surrounding a single triangle. So we conclude that $F_{p}$
are special cases of closed string operators. This observation gives
physical meaning to the above wave function. We expand the wave
function:

\begin{eqnarray}
|\psi_{\lambda}\rangle =(1 +
\sum_{p}\lambda_{p}F_{p}+\sum_{p_{1},p_{2}}\lambda_{p_{1}}\lambda_{p_{2}}F_{p_{1}}F_{p_{2}}
+ \cdots)|\psi_{0}\rangle
\end{eqnarray}
each term is a closed string state. So $|\psi_{\lambda}\rangle$ are
coherent states of closed strings.

We mention that there are also two types of strings in the models of
Kitaev and Wen \cite{kitaev1,wen1,wen2}, but strings in our model
have the additional feature that $T_{1}$ and $T_{2}$ strings walk on
triangles and hexagons, respectively. So their properties are
different. One of the differences is that there are twice as many
triangles as hexagons and thus more room for open $T_{1}$ string
endpoints than for $T_{2}$ string endpoints.

\emph{Exact spin correlation.} --The spin-spin correlation functions
can be extracted from the exact wave functions. Since wave functions
of all eigenstats are known, we can obtained all the spin-spin
correlations by direct computation:

\begin{eqnarray}
\sigma_{i}^{a}\sigma_{j}^{b}|\psi_{\lambda}\rangle = \prod_{s
}(1-\lambda_{s}F_{s}) \prod_{t}(1+\lambda_{t}F_{t})
\sigma_{i}^{a}\sigma_{j}^{b}|\psi_{0}\rangle
\end{eqnarray}
Where $s$ exhausts elementary plaquettes with the property
$\{P_{s},\sigma_{i}^{a}\sigma_{j}^{b}\}=0$, and $t$ exhausts
elementary plaquettes with the property
$[P_{s},\sigma_{i}^{a}\sigma_{j}^{b}]=0$.
 We can check that such $s$ exist except in the trivial cases when $i=j$
 and $a=b$. Since $
 (1+\lambda_{s}F_{s})(1-\lambda_{s}F_{s})=0  $, we obtain the exact
 result:
\begin{eqnarray}
\langle\psi_{\lambda}|\sigma_{i}^{a}\sigma_{j}^{b}|\psi_{\lambda}\rangle
= 0
\end{eqnarray}
except the trivial cases $
\langle\psi_{\lambda}|\sigma_{i}^{a}\sigma_{i}^{a}|\psi_{\lambda}\rangle
= \langle\psi_{\lambda}|1|\psi_{\lambda}\rangle = 1$. This
computation is analogous to the computation of spin-spin
correlations in the Kitaev model \cite{baskaran}, where they are
found to be exactly zero beyond the nearest neighbor. The spin-spin
correlations in our model are even more exotic. They are identically
zero for two different sites including the case of nearest
neighbors. We also mention that the spin correlations in the earlier
models \cite{wen1,kitaev1} can also be computed using explicit
wavefucntions. Our method can also be generalized to computation of
multi-spin correlations. Another feature is that the result is valid
for all energy eigenstates, besides the ground state. This
remarkable feature is also shared by the Kitaev's model on honeycomb
lattice \cite{kitaev2, baskaran}.

The zero spin correlations indicate very strong quantum fluctuation
in our model. So the conventional idea of using spins as local order
parameter will not give meaningful result here. But there is
topological order \cite{wen1} in this model manifesting in ground
state degeneracy on closed manifold and gapless edge states at the
boundary of open systems.

\emph{Anyonic excitation.} --The anyonic excitation in our model has
similar properties to earlier models \cite{kitaev1,kitaev2,wen2}.
The energy of an open string is not changed if we move the two
endpoints far apart, i.e. the endpoints are deconfined. We call
these endpoints $T_{1}$ and $T_{2}$ vortexes. We also mention that
these vortexes can be interpreted as $Z_{2}$ vortex excitations of
an emergent $Z_{2}$ gauge field \cite{wen1,senthil}.

The vortex hopping operators is very simple, they are the same
matrices appearing in the definition of $P_{t}$. The statistical
angles of $T_{1}$ and $T_{2}$ vortexes can be shown to be both zero
by the statistical algebra method \cite{levin}. So the same type of
vortexes see each other as bosons. The mutual statistical angle of
$T_{1}$ and $T_{2}$ vortexes can be computed in a different way.
Following Kitaev's work \cite{kitaev1}, we consider the physical
process of moving $T_{2}$ vortex $a$ around $T_{1}$ vortex $b$ along
the path of string $u$ (shown in Fig.3). The initial state is
$|\psi_{i}\rangle =P_{s}P_{t}|\psi\rangle$, where $|\psi\rangle$ is
a state with all vortexes far away from $a,b$. The final state is:

\begin{eqnarray}
 |\psi_{f}\rangle = P_{u}|\psi_{i}\rangle =
 P_{u}P_{s}P_{t}|\psi\rangle =
 -P_{s}P_{t}P_{u}|\psi\rangle =
-|\psi_{i}\rangle
\end{eqnarray}
the result does not depends on detail shape of $u$. The only
important thing is that $b$ is enclosed by $u$. So the phase factor
$-1$ is purely topological. The mutual statistical angle $\theta$ is
defined by $e^{2i\theta}=-1$. So we have $\theta=\pi /2$ which is
half of fermion's. This anyonic mutual statistic is a signal of
topological order in our model.

\begin{figure}
\includegraphics[width=8.5cm, height=6.0cm]{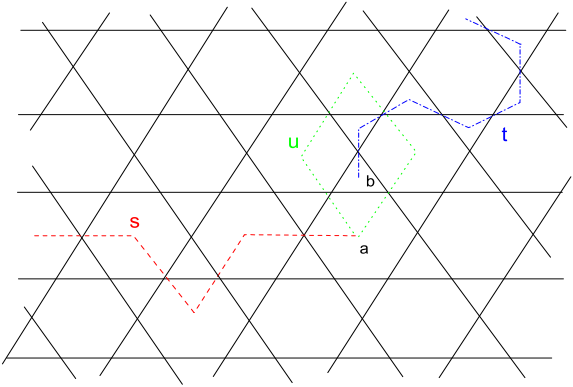}
\caption{Statistic of open string endpoints. $s$ is an open $T_{2}$
string with one endpoint $a$ and the other endpoint far away, while
$t$ is an open $T_{1}$ string with one endpoint $b$ and the other
endpoint also far away. $u$ is a closed string enclosing $b$.}
\label{fig.3}
\end{figure}

\emph{Ground state degeneracy and gapless edge states.} --The ground
state degeneracy on a closed manifold often indicate topological
order hidden in a phase \cite{wen3,wen1,niu}. To study the
topological order of our model, we consider a periodic system, i.e.
we put the system on a torus. For a $m\times n$ system (as Fig.4),
there are $2mn$ triangles, $mn$ hexagons, and $3mn$ sites (the sites
at opposite boundaries are actually the same site and should not be
re-counted). Now the mutual independency of $F_{p}$ is destroyed by
the periodic boundary condition. Actually, one can readily check
that $ \prod_{p \in triangle}F_{p}=1$ and $ \prod_{p \in
hexagon}F_{p}=1$, so the number of independent $F_{p}$ are $3mn-2$.

It has been mentioned that closed string operators $P_{t}$ commute
with $H$ and thus do not change the energy eigenvalue. But does
$P_{t}$ have chance to transform a given energy eigenstate to
another eigenstate that degenerate with it? For the cases when $t$
is homotopically trivial, i.e. $t$ is able to be deformed to a
point, the answer is no. Actually, when $t$ is a homotopically
trivial $T_{1}$ string, it can be checked that
$P_{t}=\prod_{p}F_{p}$, with $p$ exhausting the hexagons enclosed by
$t$. So $P_{t}$ cannot transform a given energy eigenstate to
another. Similar result holds for $T_{2}$ strings.

But there are also homotopically nontrivial strings called global
strings on the torus. Consider four global strings $t_{1}$, $t_{2}$,
$t_{3}$ and $t_{4}$ shown in Fig.4. We have $\{P_{t_{1}},
P_{t_{3}}\}=0$ and $\{P_{t_{2}}, P_{t_{4}}\}=0$ by their
intersecting times. All other pairs of string operators are
commutative. Since all these operators commute with $F_{p}$, we can
choose the common eigenstates of $P_{t_{1}}$, $P_{t_{2}}$ and
$F_{p}$ as the basis of the Hilbert space. To find whether
eigenvalues of $P_{t_{1}}$, $P_{t_{2}}$ and $F_{p}$  can label all
states completely, we just need to do a simple counting: All of
$P_{t_{1}}$, $P_{t_{2}}$ and $F_{p}$ have two eigenvalues $\pm1$, so
they can label $2^{3mn-2}\times2\times2=2^{3mn }$ states. This is
just the dimension of physical Hilbert space. So $P_{t_{1}}$,
$P_{t_{2}}$ and $F_{p}$ are complete labels and we have found all
energy eigenstates on the torus. From the above discussion we
conclude that states with given $F_{p}$ has additional 4-fold
degeneracy coming from two global string operators $P_{t_{1}}$ and
$P_{t_{2}}$. It follows from this that states with all $F_{p}=1$ are
4-fold degenerate, i.e. the ground state degeneracy is $4$.

\begin{figure}
\includegraphics[width=8.5cm, height=6.0cm]{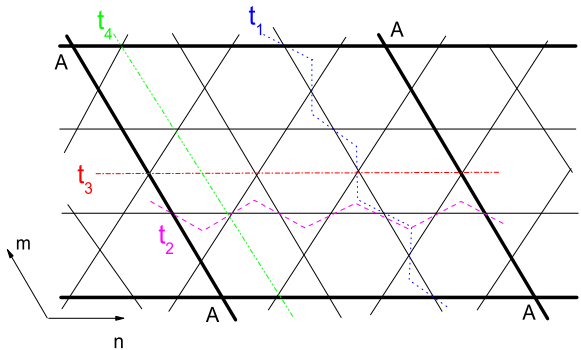}
\caption{Kagome lattice on torus. $m=n=3$ in this figure. The
effective region is within the dark bold lines. The four copies of
the point ``$A$'' should be regarded as one and the same point
because of periodic boundary condition. $t_{1}$, $t_{2}$, $t_{3}$
and $t_{4}$ are all closed strings on the torus. $t_{1}$, $t_{2}$
are $T_{1}$ strings, while $t_{3}$, $t_{4}$ are $T_{2}$ strings.
None of them can be deformed to a point and thus they are called
global strings.} \label{fig.4}
\end{figure}

 Furthermore, we can figure out how these
eigenstates transform under $P_{t_{3}}$ and $P_{t_{4}}$. Since
$\{P_{t_{3}}, P_{t_{1}}\}=0$ and $[P_{t_{3}}, P_{t_{2}}]=0$, we
conclude that $P_{t_{3}}$ change the sign of $P_{t_{1}}$ but not of
$P_{t_{2}}$. By similar argument, we can also show that $P_{t_{4}}$
change the sign of $P_{t_{2}}$ but not of $P_{t_{1}}$. This argument
also shows that both $\pm1$ eigenvalues of $P_{t_{1}}$ and
$P_{t_{2}}$ are accessible by some physical states.

We mention that $t_{1}$ and $t_{2}$ can be chosen rather
arbitrarily, with only requirements that they circle the torus in
$m$ and $n$ directions respectively, and they are of the same string
type. The physics here will not change with our choices because it
is topological.

We also find the existence of gapless edge states in our model,
which is another manifestation of topological order. If we cut the
torus open along the bold lines in Fig.4, the dimension of Hilbert
space is larger by a factor $2^{L}$ than the case of torus, where
$L\approx m+n$ when $m,n$ are large. But the number of $F_{p}$
labeling energis is unchanged when we cut the torus open. So there
is a large degeneracy proportional to $2^{L}$, which we can
interpret as various edge excitations with zero energy
\cite{wen1,wen2}.

\emph{Conclusion.} --An exact spin liquid state has been given on
the kagome lattice based on multi-spin interactions. All spin-spin
correlations are found to be zero except trivial cases. There are
also anyonic excitations and ground state degeneracy indicating
nontrivial topological order. We think that this exact spin liquid
state with odd number of spins per unit cell might be helpful to
improve our understanding of $2$D spin liquid states. Furthermore,
the existence of such exact states suggests that there maybe exist
other interesting states based on multi-spin interactions.

The work was supported by National Natural Science Foundation of
China under Grant No. 10675108.


\begin{thebibliography}{99}

\bibitem{wen1}  X.G. Wen, \emph{Quantum Field Theory of Many-Body
Systems} (Oxford University, New York, 2004).
\bibitem{nayak} C. Nayak, S. Simon, A. Stern, M. Freedman and S.
Sarma, Rev. Mod. Phys. 80, 1083 (2008).
\bibitem{sachdev} S. Sachdev, \emph{Quantum Phase Transitions}
(Cambridge University Press, Cambridge, England, 1999).
\bibitem{laughlin} R. B. Laughlin, Phys. Rev. Lett. 50, 1395 (1983).
\bibitem{anderson} G. Baskaran and P. W. Anderson, Phys. Rev. B 37, 580
(1988).
\bibitem{kitaev1}  A. Kitaev, Ann. Phys. 303, 2 (2003).
\bibitem{kitaev2}  A. Kitaev, Ann. Phys. 321, 2 (2006).
\bibitem{wen2} X.G. Wen, Phys. Rev. Lett. 90, 016803 (2003).
\bibitem{yao}  H. Yao and S. Kivelson, Phys.
Rev. Lett. 99, 247203 (2007).
\bibitem{schroeter} D.F. Schroeter, E. Kapit, R. Thomale, and M.
Greiter, Phys. Rev. Lett. 99, 097202 (2007).
\bibitem{sun} S. Yang, D.L. Zhou and C.P. Sun, Phys. Rev. B 76, 180404(R)
(2007).
\bibitem{xiang} X.Y. Feng, G.M. Zhang and T. Xiang, Phys. Rev. Lett.
98, 087204 (2007).
\bibitem{kalmeyer} V. Kalmeyer and R. B. Laughlin, Phys. Rev. Lett. 59, 2095 (1987).
\bibitem{baskaran} G. Baskaran, S. Mandal and R. Shankar, Phys. Rev.
Lett. 98, 247201 (2007).
\bibitem{senthil} T. Senthil and M. Fisher, Phys. Rev. B 62, 7850
(2000).
\bibitem{levin} M. Levin and X.G.Wen, Phys. Rev. B 67, 245316
(2003).
\bibitem{wen3} X.G. Wen, Phys. Rev. B 40, 7387 (1989).
\bibitem{niu} X.G. Wen and Q. Niu, Phys. Rev. B 41, 9377 (1990).

\end{thebibliography}
\end{document}